\documentclass[prl,twocolumn,showpacs,preprintnumbers,amsmath,amssymb]{revtex4-1}
\usepackage{graphicx}% Include figure files
\usepackage{dcolumn}% Align table columns on decimal point
\usepackage{bm}% bold math

\begin{document}

\title{Rotating Dipolar Spin-1 Bose-Einstein Condensates}
\author{T.~P. Simula$^{1,2}$, J.~A.~M. Huhtam\"aki$^{2,3}$, M. Takahashi$^{2,4}$, T. Mizushima$^2$, and K. Machida$^2$}
\affiliation{$^1$School of Physics, Monash University, Victoria 3800, Australia\\$^2$Department of Physics, Okayama University, Okayama 700-8530, Japan\\ $^3$Department of Applied Physics/COMP, Aalto University School of Science and Technology, P.O.Box 15100, FI-00076 AALTO, Finland\\$^4$Department of Physics, The University of Tokyo, Tokyo 113-0033, Japan}

\begin{abstract}
We have computed phase diagrams for rotating spin-1 Bose-Einstein condensates with long-range magnetic dipole-dipole interactions. Spin textures including vortex sheets, staggered half-quantum- and skyrmion vortex lattices and higher order topological defects have been found. These systems exhibit both superfluidity and magnetic crystalline ordering and they could be realized experimentally by imparting angular momentum in the condensate.

\end{abstract}

\maketitle

% Intro ================================================================================
A bar magnet is an omnipresent dipolar object in classical physics. By dislodging one magnet in the proximity of another the striking dependence of the interaction between dipoles on their separation and relative orientation becomes apparent. In addition to magnetism, dipolar effects play role in a multitude of physical systems and processes including wireless telecommunications, spintronics, chemical bonding between atoms and molecules, and liquid crystals \cite{Gennes1993a}.

In the recent years, several candidates exhibiting dipolar effects in ultra-cold quantum gas experiments have emerged \cite{Lahaye2009a}. Polar molecules with permanent electric dipole moments \cite{Ospelkaus2008a} may provide a system amenable to a semiclassical treatment \cite{Takahashi2007a,Huhtamaki2010a}. Chromium condensates have been observed to show dipolar effects due to their intrinsically large magnetic moments \cite{Stuhler2005a,Lahaye2008a} with dysprosium \cite{Lu2010a} and erbium promising even stronger dipolar effects and elevated potential for quantum simulators and quantum information processing applications \cite{Lahaye2009a}. It may also be possible to achieve co-existence of both superfluid and magnetic long-range order in these systems and to realize elusive supersolid states---hints of which has already been observed in experiments with spinor $^{87}$Rb \cite{Vengalattore2008a,Vengalattore2009a}. In addition, dipolar cold atomic clouds are predicted to feature a range of quantum phenomena previously unseen in Bose-Einstein condensed gases \cite{Santos2003a,Kawaguchi2006b,Kawaguchi2006a,Santos2006a}.

Here we calculate phase diagrams for rotating spin-1 Bose gas in the presence of long-range magnetic dipole-dipole interactions without constraining the spin degree of freedom. On changing the relative strength of the long-range dipole-dipole interaction with respect to the local contact interaction, structural lattice transitions occur in the system. Due to the dipolar interactions a variety of textures including vortex sheets, staggered skyrmion and half-quantum vortex lattices and higher order topological defects emerge in these systems. 

% Model: GP ==========================================================
\emph{Model---}
We consider a zero-temperature spin-1 Bose-Einstein condensate and model it using a mean-field theory which includes magnetic dipole-dipole interactions. Such system is described by a three-component spinor order parameter  $\Psi=[\psi_{1}({\bf r}),\psi_0({\bf r}),\psi_{-1}({\bf r})]^T$ \cite{Ohmi1998a,Ho1998a}. The ground-state wavefunction is a solution to a dipolar spinor Gross-Pitaevskii equation
\begin{equation}
\left( {\begin{array}{ccc}
 \mathcal{L} +f_z +d_z  &  f_{xy}^*  +d^*_{xy} & 0 \\
f_{xy}  +d_{xy}                   & \mathcal{L}      &  f_{xy}^*  +d^*_{xy} \\
0                                       & f_{xy}  +d_{xy}     &  \mathcal{L}  -f_z - d_z
 \end{array} } \right)
\left( {\begin{array}{c}
 \psi_1\\
 \psi_0 \\
 \psi_{-1}
 \end{array} } \right)
=0, 
\label{GP}
\end{equation}
where the operator
$\mathcal{L}({\bf r}) = -\hbar^2\nabla^2/2m+V_{\rm ext}({\bf r}) - \Omega L_z -\mu +gn({\bf r})$
contains a single particle part and a contact interaction term, where $n({\bf r})=\sum_k|\psi_k({\bf r})|^2$ is the total particle density. Notwithstanding there does not exist, due to dipolar interparticle potential, a rotating reference frame where the laws of equilibrium statistical mechanics would apply \cite{Leggett2008a}, we may assign $\Omega$ to be a Lagrange multiplier associated with the angular momentum of the system measured by the operator $L_z$. The external potential $V_{\rm ext}({\bf r})=m [\omega^2_\perp(x^2+y^2) + \omega_z^2z^2]/2$ is fixed by the transverse $\omega_\perp$ and axial $\omega_z$ harmonic frequencies. The normalization condition of the wavefunction, $\int_Vn({\bf r}) d{\textbf r}=N$, relates the number of particles $N$ in the system to the chemical potential $\mu$. The spin-exchange interaction yields the terms
$f_z({\bf r}) = g_s \langle F_z\rangle$
and
$f_{xy}({\bf r})=g_s(\langle F_x\rangle+i\langle F_y\rangle)/\sqrt{2}$.
The terms
$d_z({\bf r}) =g_d I_z$
and
$d_{xy}({\bf r}) =g_d(I_x+iI_y)/\sqrt{2}$,
where
$I_k({\bf r}) = \int d{\bf r'} (\langle F_k\rangle-3e_k \sum_j e_j\langle F_j\rangle)/ |{\bf r}-{\bf r'}|^3)$
are due to the dipole-dipole interaction. Here the vector ${\bf e}=({\bf r}-{\bf r'})/|{\bf r}-{\bf r'}|$ and $F_k$ denote the components of the spin-1 operator ${\bf F}$ and $\langle\cdot\rangle$ are their expectation values in the ground state. The coupling constants $g=4\pi\hbar^2 (a_0+2a_2)/3m, g_s=4\pi\hbar^2 (a_2 - a_0)/3m$ and $g_d=\mu_0\mu^2_{\rm B}g^2_F/4\pi$ determine, respectively, the strength of $s$-wave, spin-exhange and dipole-dipole interactions between particles of mass $m$ in terms of the bare scattering lengths $a_0$ and $a_2$ and the permeability of vacuum $\mu_0$, Bohr magneton $\mu_{\rm B}$ and the Land\'e factor $g_F$.  

We use $\hbar\omega_\perp$ and $a_\perp=\sqrt{\hbar/m\omega_\perp}$ for the units of energy and length, respectively. Our computational system is specified by the aspect ratio $\omega_z=10 \;\omega_\perp$ and dimensionless parameters $gN/\hbar\omega_\perp a_\perp^3=1000$, $g'_s=g_s/g$ and $g'_d=g_d/g$. As in \cite{Yi2006a}, we choose $g'_s=0.03$ for antiferromagnetic and $g'_s=-0.01$ for ferromagnetic spin-exchange interaction corresponding to the values for $^{23}$Na and $^{87}$Rb, respectively. The dimensionless frequency $\Omega'=\Omega/\omega_\perp$ and $g'_d$ are the adjustable parameters of our numerical experiments. We assume that external magnetic fields are absent and therefore exclude linear and quadratic Zeeman terms from Eq.(\ref{GP}). 

% FIGURE ===
\begin{figure}
\includegraphics[width=\columnwidth]{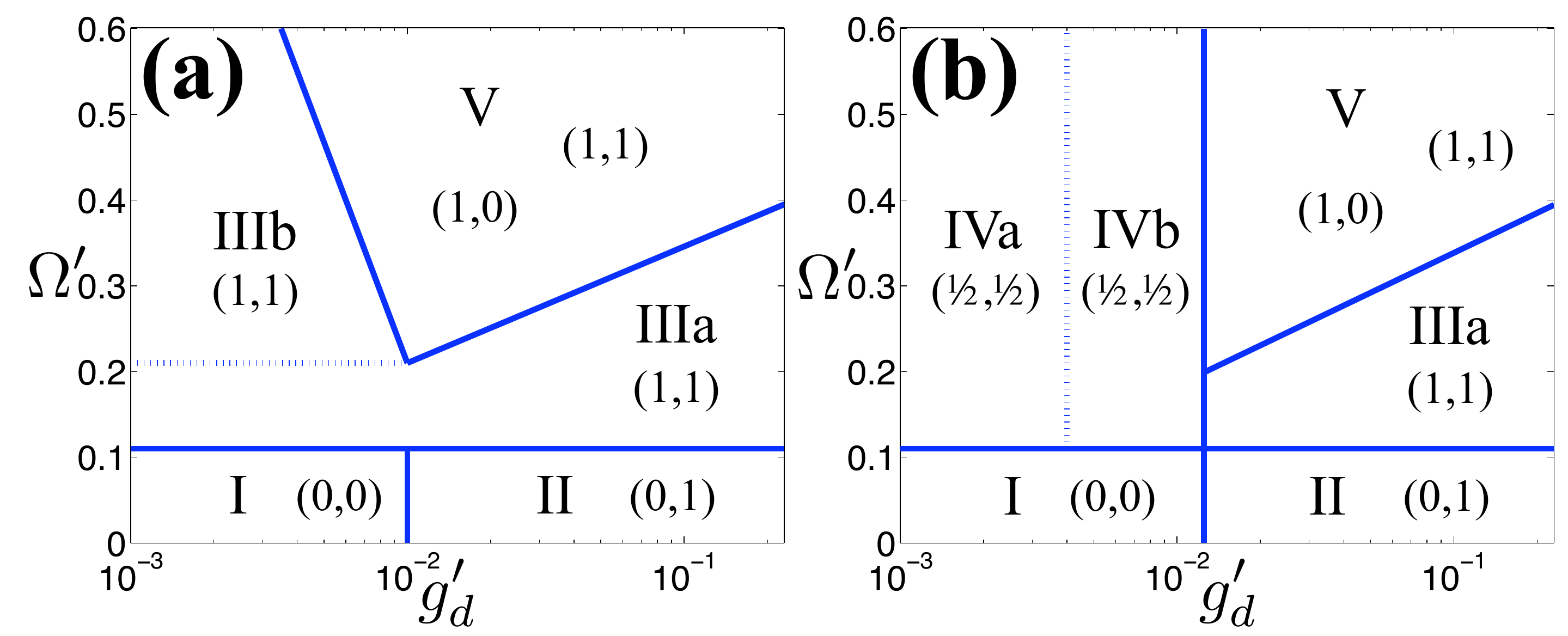}
\caption{Schematic phase diagrams in the $g_d'-\Omega'$ parameter space for (a) ferromagnetic $g_s'=-0.01$ and (b) antiferromagnetic $g_s'=0.03$ spin-exchange interaction. The numbers denote the type of topological defects in each phase as described in the text. Phase boundaries are estimated by straight lines.}
\label{pd}
\end{figure}
% FIGURE ===

% VORTEX TYPES ===============================================
\emph{Elementary vortex types---}
To facilitate further discussion we categorize certain topological defects relevant to our system. Invariance of the spinor order parameter $\Psi$ under a suitable combination of spin rotation and gauge transformation $\Psi'=e^{i\theta} e^{-i\phi{\bf n}\cdot{\bf F}}\Psi$, where ${\bf n}$ is a unit vector along the spin-rotation axis, allows us to classify the line defects in our system by associating with them a pair of winding numbers $(\ell,s)$. Here $\ell=|\theta| / 2\pi$ measures the strength of mass circulation about the defect line and $s=|\phi| /2\pi$ is the spin winding number. With this notation, a vortex which carries one unit of orbital angular momentum per particle and involves a $2\pi$ spin rotation is denoted by $(1,1)$. Such vortex has a ferromagnetic core and depending on the underlying model and imposed boundary conditions is known as a Mermin-Ho- \cite{Mermin1976a} or Anderson-Toulouse vortex, \cite{Anderson1977a} or a skyrmion \cite{Skyrme1961a}. These have recently been created  in transition metal compound MnSi \cite{Neubauer2009a} and in BECs as dynamical states \cite{Leslie2009a}.

A half-quantum vortex or an ``Alice string" is characterized by the doublet ($\slantfrac{1}{2},\slantfrac{1}{2}$) and is predicted to exist e.g. in superconductors, superfluid helium systems and spinor BECs \cite{Volovik2003a,Isoshima2002a,Ruostekoski2003a}. The half-quantum vortex has a ferromagnetic core structure and vanishing spin density elsewhere. Half-quantum vortices have recently been created in polariton condensates \cite{Lagoudakis2009a}.

In addition to the two ``mixed" spin--mass textures, pure spin vortices (0,1)  and pure mass vortices (1,0) emerge in our system. The former has a core devoid of spins and is therefore said to possess an unmagnetized polar core, whereas the spin currents vanish around the latter. All of the above described topological objects are``coreless vortices" in the sense that the total particle density is finite at the vortex cores \cite{Nootti}.

% FIGURE ===
\begin{figure}
\includegraphics[width=\columnwidth]{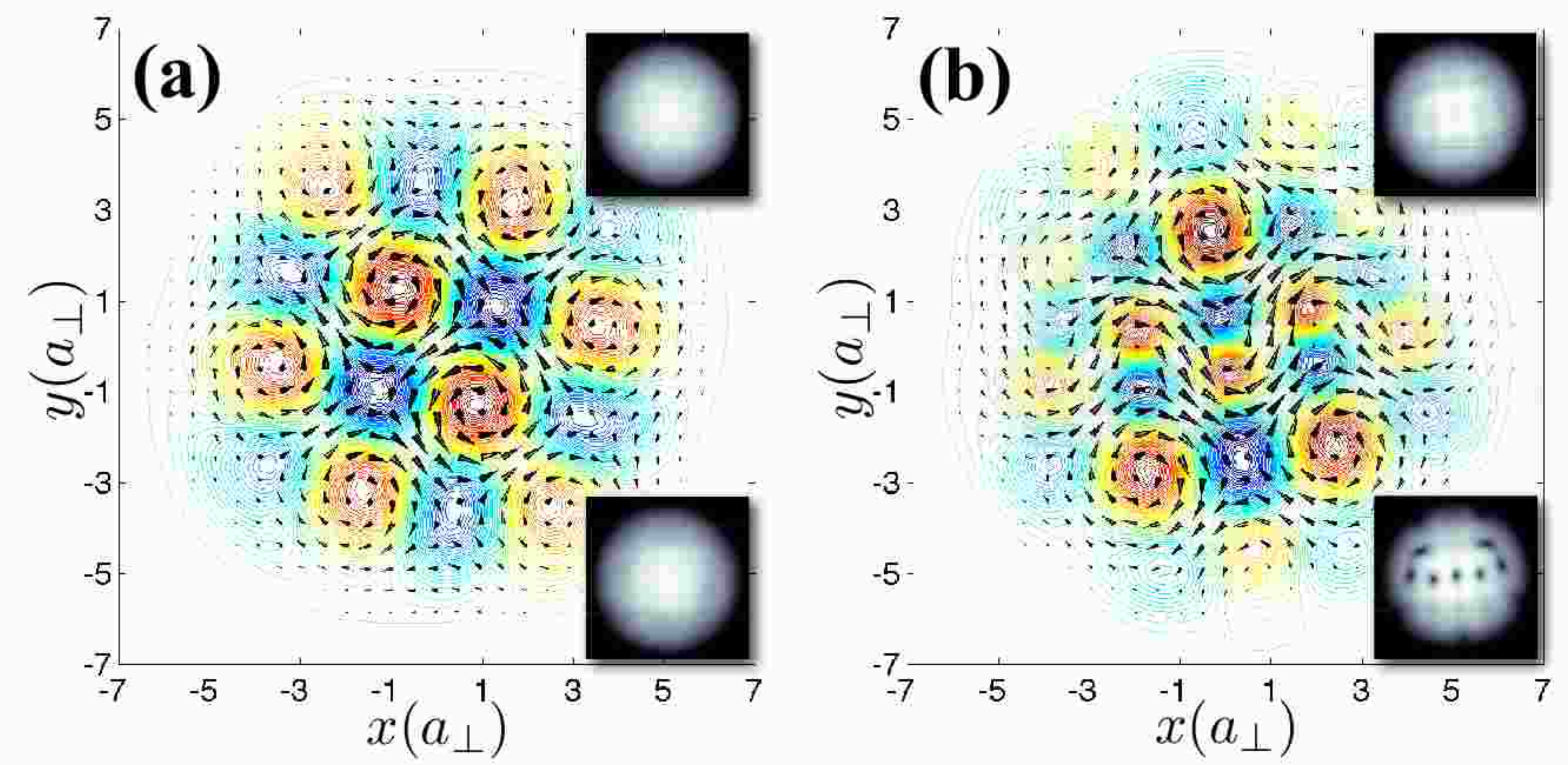}
\caption{(Color online) Skyrmion textures for $g'_s=-0.01$, $\Omega'=0.6$ and (a) $g'_d=0.001$ and (b) $g'_d=0.01$ corresponding to phases IIIb and V, respectively. The main frames show the spin texture, arrows indicating the orientation of local spin and contour lines are plotted for the $z$-component of spin density. Insets to the right of the main frames show the particle density $n({\bf r})$ and spin density $\sqrt{\sum_k\langle F_k\rangle^2}$.  All square frames have the same ($7a_\perp\times7a_\perp$) spatial extent.} 
\label{ferro}
\end{figure}
% FIGURE ===

% PHASE DIAGRAM  FERRO =================================================
\emph{Ferromagnetic spin-exchange interaction---}
Figure \ref{pd} (a) shows a schematic phase diagram in the $g_d'-\Omega'$ parameter space for a ferromagnetic $g'_s=-0.01$ spin-exchange interaction. For $\Omega' = 0$ two phases exist depending on the relative strength of the dipolar interactions \cite{Yi2006a,Kawaguchi2006b,Takahashi2007a,Huhtamaki2010a}. In I the ground states are free of topological defects whereas in II spontaneous spin currents emerge in the system. The ground states in II are occupied by a single (0,1) vortex \cite{Yi2006a}. Eventually, for $g_d'\approx 1/4$ the system becomes unstable due to the effective attraction caused by the dipole-dipole term \cite{Yi2006a,Huhtamaki2010a,Parker2009a,Lahaye2009a}. The critical value of $g_d'$ is fairly insensitive to the values of $\Omega'$ and $g_s'$ for the range of parameters studied here. In experiments, the value of $g_d'$ has been successfully tuned for spin-polarized chromium condensates \cite{Lahaye2008a}. In the absence of magnetic fields this could be achieved using optical Feshbach resonances \cite{Theis2004a}.

For $g_d'=0$ on increasing $\Omega'$ above a critical value, mass currents are nucleated in the system \cite{Mizushima2004a}. For ferromagnetic spin-exchange term $g_s'<0$ the ground states in IIIa contain a single $(1,1)$ vortex. The phase IIIa precedes the dipolar collapse whenever $\Omega'$ exceeds the critical value. For small values of $g_d'$ on increasing $\Omega'$ further staggered skyrmion lattices composed of ($1,1$) vortices form as illustrated in Fig. \ref{ferro}(a) which displays a typical phase IIIb texture. The arrows in the main frame show the local spin orientation in the $z=0$ plane. The contour lines are plotted for the $z$-component of the local spin density. The inset in the top right corner shows the particle density. Due to the weak dipolar coupling and since the skyrmion cores are filled with spins which are oriented parallel and antiparallel to the $z$-axis, the spin density shown in the bottom right corner is rather smooth. The spin-texture in Fig. \ref{ferro}(a) is seen to be composed of two interleaved skyrmion lattices of opposite vortex core polarisations and spin windings. Generally, individual $(1,1)$ vortices do not possess cylindrical symmetry in these systems. On further increasing the value of $g_d'$ keeping $\Omega'$ fixed, type ($1,0$) vortices enter the system \cite{Mizushima2002a}. A variety of textures composed of a mixture of ($1,1$) and ($1,0$) vortex molecules are observed in phase V, an example of which is illustrated in Fig. \ref{ferro} (b). The ($1,0$) vortices leave a clear signature in the spin density due to their spinless core structure. 

\begin{figure}
\includegraphics[width=\columnwidth]{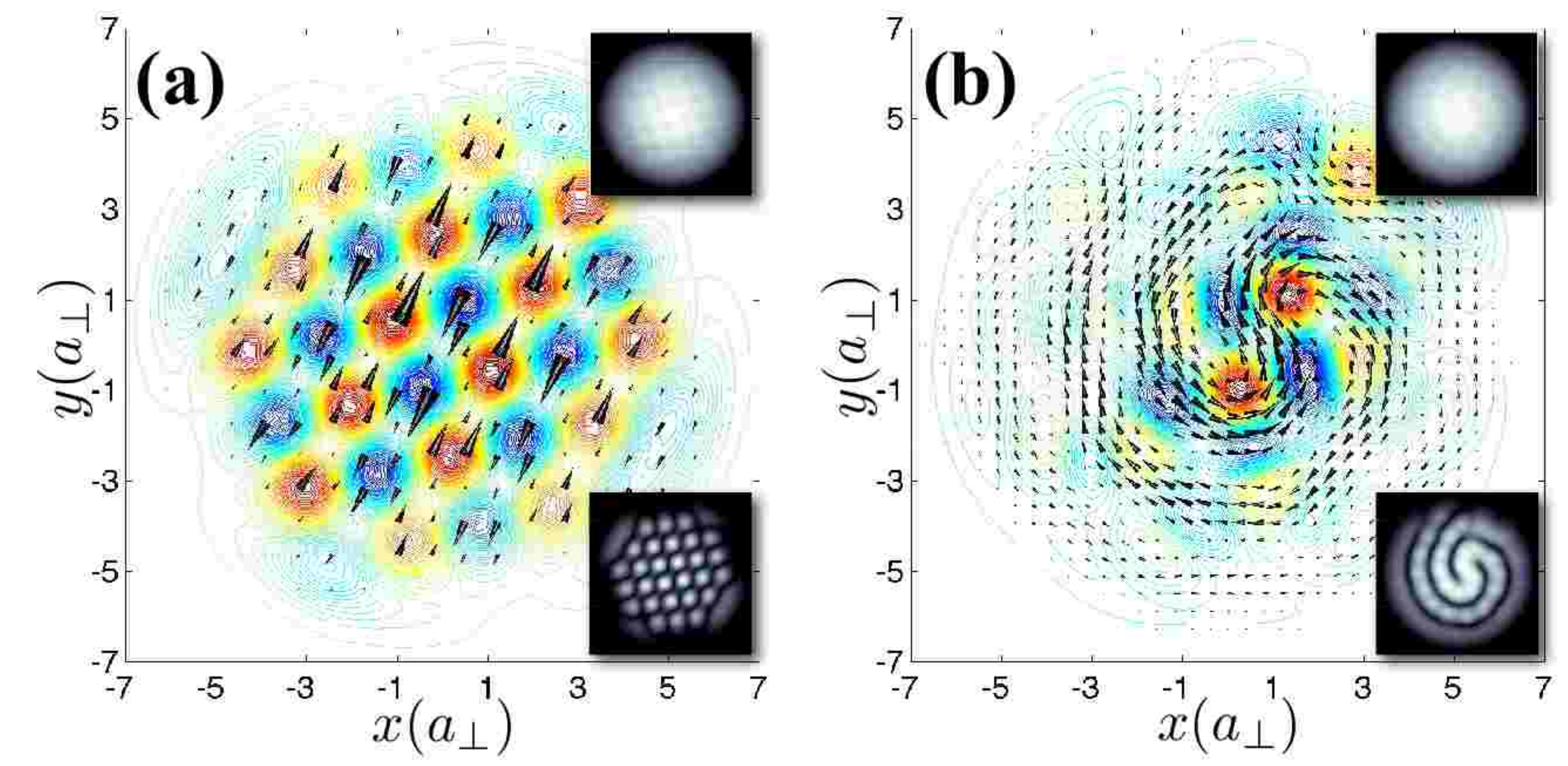}
\caption{(Color online) Magnetic crystal and vortex sheet textures respectively for (a) $g'_d=0.001$ and (b) $g'_d=0.01$ corresponding to phases IVa and IVb. Contour lines in (a) and (b) are plotted for the $x$- and $z$-components of spin density, respectively. For both (a) and (b) $g'_s=0.03$ and $\Omega'=0.6$. All square frames have the same ($7a_\perp\times7a_\perp$) spatial extent.} 
\label{anti}
\end{figure}

% PHASE DIAGRAM  ANTIFERRO =================================================
\emph{Antiferromagnetic spin-exchange interaction---}
For antiferromagnetic spin-exchange term $g_s'>0$ the phase diagram shown in Fig.\ref{pd}(a) reduces to that of the ferromagnetic one shown in Fig.\ref{pd}(b) both for small $\Omega'$ and for large $g_d'$ where dipolar effects dominate over the spin-spin interaction. In phase IVa the constituent defects  are of type ($\slantfrac{1}{2},\slantfrac{1}{2}$) and their spins lie predominantly in the $x$-$y$ plane. The spins in the cores of the nearest neighbor defects are pointing in the opposite directions forming a staggered antiferromagnetic structure as shown in Fig.\ref{anti}(a). An image of the spin density strikingly reveals the underlying magnetic crystal. Magnetization density in these rotating states is analogous to the fractional $1/3-$vortex lattice states found in the cyclic state of $F=2$ spinor condensates \cite{Huhtamaki2009a}. On crossing from IVa to IVb the square lattice of ($\slantfrac{1}{2},\slantfrac{1}{2}$) vortices is first squeezed into a rhombic shape due to the head-to-tail attraction of the dipoles. Next the staggered magnetic crystal phase IVa melts and the vortices rearrange to form a vortex sheet in phase IVb. In a vortex sheet the spin axes of the ($\slantfrac{1}{2},\slantfrac{1}{2}$) vortices tend to align with the orbital angular momentum. Figure \ref{anti} (b) illustrates a typical texture emerging in phase IVb. Similar vortex sheets were first observed in superfluid $^3$He-$A$ \cite{Parts1994a}. For even larger values of $g_d'$, in phase V, the ($\slantfrac{1}{2},\slantfrac{1}{2}$) vortices are converted to ($1,1$) and ($1,0$) defects where the former (latter) can be thought to be constructed by merging two ($\slantfrac{1}{2},\slantfrac{1}{2}$) vortices with same (opposite) spin windings. In the same spirit, the ($0,1$) polar core vortices cannot be built from ($\slantfrac{1}{2},\slantfrac{1}{2}$) vortices since the external rotation breaks the time reversal symmetry.

\begin{figure}
\includegraphics[width=\columnwidth]{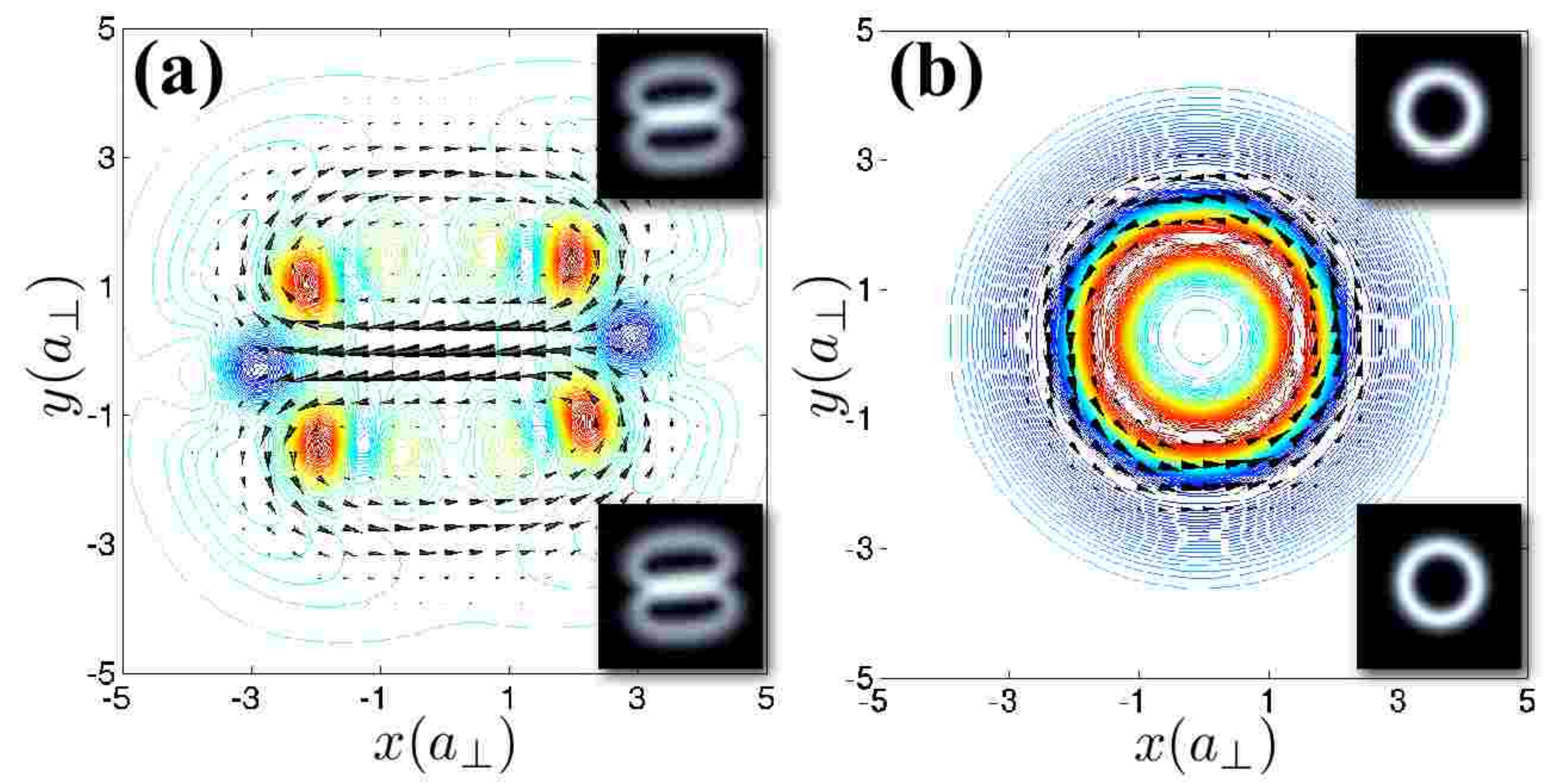}
\caption{(Color online) Higher order topological defects calculated for (a) $g_d'=0.23$ and (b) $g_d'=0.24$. For both states $g_s'=-0.01$, $\Omega'=0.95$, contour lines are plotted for the $z$-component of spin density and all square frames have ($5a_\perp\times5a_\perp$) spatial extent. } 
\label{butt}
\end{figure}

% HIGHER ORDER DEFECTS =================================================
\emph{Higher order topological defects---}
Rapid rotation of the system leads to a centrifugal explosion of the cloud whereas strong enough dipole-dipole interaction causes the condensate to implode. Therefore the simultaneous limit $\Omega' \to 1$ and $g_d'\to 1/4$ deserves particular attention. Approaching this limit, Fig.\ref{butt} shows textures under rapid rotation in strongly dipolar systems. In Fig.\ref{butt} (a), we interpret each loop of the `figure of eight' as deformed type $(4,1)$ defect and in Fig.\ref{butt} (b) a single $(5,1)$ defect is identified. In addition, we have found a wide variety of both axisymmetric and non-axisymmetric high winding number defects. It is thus conceivable that a whole hierarchy of higher order topological defects emerge in this dipolar system. In the absence of dipolar forces, higher order topological objects are suppressed due to the quadratic increase of the kinetic energy as a function of the multiplicity of the topological defect. Here the addition of dipolar interactions enables the existence of such textures irrespective of the sign of the spin-spin coupling. Furthermore, due to the intrinsic effective confinement provided by the dipolar interactions which, together with the trapping potential, may compensate for the centrifugal expansion, the experimental realization of the lowest Landau level physics may become reachable in these rapidly rotating dipolar systems.

% FIGURE ===
% FIGURE ===

% FIGURE ===
% FIGURE ===

% Conclusions ================================================================================
In conclusion, we have calculated phase diagrams for rotating dipolar $F=1$ spinor Bose-Einstein condensates. For weak dipolar coupling and for ferromagnetic (antiferromagnetic) spin-exchange interaction the spin textures composed of skyrmions (half-quantum vortices) arrange in an interlaced double square lattice structures where the magnetized vortex cores exhibit staggered antiferromagnetic ordering. The orbital angular momentum loaded in the system causes quantized mass currents to nucleate in the condensate due to its inherent superfluid property. As a consequence magnetic crystalline ordering is induced in the system by the spin vortices generated via the spin-gauge symmetry. These steadily rotating systems are superfluid and form equilibrium magnetic crystals. They also exhibit spatial periodic modulation in the particle density. For antiferromagnetic spin-exchange coupling on increasing the dipolar interaction strength the magnetic crystals melt and form vortex sheet textures. Various spin structures emerge already for small values of relative dipolar interaction strength and might therefore be observable experimentally. Our results, albeit calculated for rotating systems, may also shed light to the interpretation of the short-range magnetic structures observed in \cite{Vengalattore2008a}.

Lattice defects such as interstitials and vacancies frequently present in the non-ground states leave an extremely clear signature in the particle density profiles and could be detected using Stern-Gerlach type experiments in combination with standard imaging techniques. Therefore these systems might also provide an excellent platform to investigate proliferation of defects in the finite-temperature Berezinskii-Kosterlitz-Thouless superfluid transition \cite{Hadzibabic2006a,Clade2007a,Simula2006a,Simula2008b,Bisset2008a,Holzmann2007a,Pietila2009a}. 

For strong enough dipole-dipole coupling the system collapses due to the induced effective attraction whereas rotation exceeding the harmonic trapping frequency leads to centrifugal explosion. This suggests the possibility of entering exotic quantum-Hall like states for which the dipolar attraction and centrifugal effect would be counter-balanced. Approaching such limit we have found a wide variety of higher order topological defects entering the system.

% References ================================================================================
\begin{acknowledgments}
This work was supported by the Japan Society for the Promotion of Science (JSPS).
\end{acknowledgments}


\begin{thebibliography}{90}
\bibitem{Gennes1993a}
P. G. de Gennes and J. Prost, \emph{The Physics of Liquid Crystals}, (Oxford University Press, Oxford 1993)
\bibitem{Lahaye2009a}
T. Lahaye, C Menotti, L. Santos, M. Lewenstein, and T. Pfau , Rep. Prog. Phys. {\bf 72}, 126401 (2009).
\bibitem{Ospelkaus2008a}
K. Ni, S. Ospelkaus, M. de Miranda, A. Pe'er, B. Neyenhuis, J. Zirbel, S. Kotochigova, P. Julienne, D. Jin, and J. Ye, Science {\bf 322}, 231 (2008). 
\bibitem{Takahashi2007a}
M. Takahashi, S. Ghosh, T. Mizushima, and K. Machida, Phys. Rev. Lett. {\bf 98}, 260403 (2007).
\bibitem{Huhtamaki2010a}
J.~A.~M. Huhtam\"aki, M. Takahashi, T. P. Simula, T. Mizushima, and K. Machida, Phys. Rev. A {\bf 81}, 063623 (2010).
\bibitem{Stuhler2005a}
J. Stuhler, A. Griesmaier, T. Koch, M. Fattori, T. Pfau, S. Giovanazzi, P. Pedri, and L. Santos, Phys. Rev. Lett. {\bf 95}, 150406 (2005).
\bibitem{Lahaye2008a}
T. Lahaye, J. Metz, B. Fr\"ohlich, T. Koch, M. Meister, A. Griesmaier, T. Pfau, H. Saito, Y. Kawaguchi, and M. Ueda, Phys. Rev. Lett. {\bf 101}, 080401 (2008).
\bibitem{Lu2010a}
M. Lu, S. H. Youn, and B. L. Lev, Phys. Rev. Lett. {\bf 104}, 063001 (2010).
\bibitem{Vengalattore2008a}
M. Vengalattore, S. R. Leslie, J. Guzman, and D. M. Stamper-Kurn, Phys. Rev. Lett. {\bf 100}, 170403 (2008).
\bibitem{Vengalattore2009a}
M. Vengalattore, J. Guzman, S. Leslie, F. Serwane, and D. M. Stamper-Kurn, arXiv:0901.3800.
\bibitem{Santos2003a}
L. Santos, G. V. Shlyapnikov, and M. Lewenstein, Phys. Rev. Lett. {90}, 250403 (2003).
\bibitem{Kawaguchi2006b}
Y. Kawaguchi, H. Saito and M. Ueda, Phys. Rev. Lett {\bf 97}, 130404 (2006).
\bibitem{Kawaguchi2006a}
Y. Kawaguchi, H. Saito and M. Ueda, Phys. Rev. Lett {\bf 96}, 080405 (2006).
\bibitem{Santos2006a}
L. Santos, T. Pfau, Phys. Rev. Lett. {96}, 190404 (2006).
\bibitem{Ohmi1998a}
T. Ohmi and K. Machida, J. Phys. Soc. Jpn. {\bf 67}, 1822 (1998).
\bibitem{Ho1998a}
T. L. Ho, Phys. Rev. Lett. {\bf 81}, 742 (1998).
\bibitem{Leggett2008a}
A. J. Leggett, \emph{Quantum Liquids} (Oxford University Press, Oxford 2006).
\bibitem{Yi2006a}
S. Yi and H. Pu, Phys. Rev. Lett. {\bf 97}, 020401 (2006).
\bibitem{Mermin1976a}
N.D. Mermin and T.-L. Ho, Phys. Rev. Lett. {\bf 36}, 594 (1976). 
\bibitem{Anderson1977a}
P. W. Anderson and G. Toulouse, Phys. Rev. Lett. {\bf 38}, 508 (1977). 
\bibitem{Skyrme1961a}
T. H. R. Skyrme, Proc. R. Soc. Lond. A {\bf 260}, 127 (1961); Nucl. Phys. {\bf 31}, 556 (1962). 
\bibitem{Neubauer2009a}
A. Neubauer, C. Pfleiderer, B. Binz, A. Rosch, R. Ritz, P.G. Niklowitz, and P. B\"oni, Phys. Rev. Lett. {\bf 102}, 186602 (2009).
\bibitem{Leslie2009a}
L. S. Leslie, A. Hansen, K. C. Wright, B. M. Deutsch, and N. P. Bigelow, Phys. Rev. Lett. {\bf 103}, 250401 (2009).
\bibitem{Volovik2003a}
G. E. Volovik, \emph{The Universe in a Helium Droplet} (Clarendon Press, Oxford, 2003).
\bibitem{Isoshima2002a}
T. Isoshima and K. Machida, Phys. Rev. A {\bf 66}, 023602 (2002).
\bibitem{Ruostekoski2003a}
J. Ruostekoski and J.R. Anglin, Phys. Rev. Lett. {\bf 91}, 190402 (2003).
\bibitem{Lagoudakis2009a}
K. G. Lagoudakis, T. Ostatnicky, A. V. Kavokin, Y. G. Rubo, R. Andr\'e B. Deveaud-Pl\'edran, Science {\bf 326}, 974 (2009).
\bibitem{Nootti}
The (1,0) vortices in our system are nonaxisymmetric ``split-core" states having finite core-density.
\bibitem{Parker2009a}
N. G. Parker, C. Ticknor, A. M. Martin, and D. H. J. O'Dell, Phys. Rev. A {\bf 79}, 013617 (2009).
\bibitem{Theis2004a}
M. Theis, G. Thalhammer, K. Winkler, M. Hellwig, G. Ruff, R. Grimm, and J. H. Denschlag, Phys. Rev. Lett. {\bf 93}, 123001 (2004).
\bibitem{Mizushima2004a}
T. Mizushima, N. Kobayashi and K. Machida, Phys. Rev. A {\bf 70}, 043613 (2004).
\bibitem{Mizushima2002a}
T. Mizushima, K. Machida and T. Kita, Phys. Rev. Lett. {\bf 89}, 030401 (2002).
\bibitem{Huhtamaki2009a}
J.~A.~M. Huhtam\"aki, T. P. Simula, M. Kobayashi, and K. Machida, Phys. Rev. A {\bf 80}, 051601(R) (2009).
\bibitem{Parts1994a}
\"U. Parts, E. V. Thuneberg, G. E. Volovik, J. H. Koivuniemi, V. M. H. Ruutu, M. Heinil\"a, J. M. Karim\"aki, and M. Krusius, Phys. Rev. Lett. {\bf 72}, 3839 (1994).
\bibitem{Hadzibabic2006a}
Z. Hadzibabic, Z. Kr\"uger, M. Cheneau, B. Battelier, and J. Dalibard, Nature {\bf 441}, 1118 (2006).
\bibitem{Clade2007a}
P. Clad\'e, C. Ryu, A. Ramanathan, K. Helmerson, and W. D. Phillips, Phys. Rev. Lett. {\rm 102}, 170401 (2009).
\bibitem{Simula2006a}
T. P. Simula and P. B. Blakie, Phys. Rev. Lett. {\bf 96}, 020404 (2006). 
\bibitem{Simula2008b}
T. P. Simula, M. J. Davis, and P. B. Blakie, Phys. Rev. A {\bf 77}, 023618 (2008).
\bibitem{Bisset2008a}
R. N. Bisset, M. J. Davis, T. P. Simula, and P. B. Blakie, Phys. Rev. A {\bf 79}, 033626 (2009).
\bibitem{Holzmann2007a}	
M. Holzmann and W. Krauth, {Phys. Rev. Lett.} {\bf 100}, 190402 (2008).
\bibitem{Pietila2009a}
V. Pietil\"a, T. P. Simula, and M. M\"ott\"onen, Phys. Rev. A {\bf 81}, 033616 (2010).
% =============================================================

\end{thebibliography}
\end{document}